\documentclass[a4paper]{jpconf}
\usepackage{graphicx}

\usepackage{bm}
\usepackage{amsmath}
\usepackage{amssymb}
\usepackage{amsfonts}
\usepackage{float}
\usepackage{dsfont}  
\usepackage{slashed}  
\usepackage{booktabs}

\newcommand{\be}{\begin{equation}}  
\newcommand{\ee}{\end{equation}}  
\newcommand{\beq}{\begin{eqnarray}}  
\newcommand{\eeq}{\end{eqnarray}}

\begin{document}
\title{Hyperon and charmed baryon masses and nucleon excited states from lattice QCD}

\author{Constantia Alexandrou}

\address{ Department of Physics, University of Cyprus, P.O. Box 20537, 1678 Nicosia, Cyprus\\  
  Computation-based Science and Technology Research  
  Center, Cyprus Institute, 20 Kavafi Str., Nicosia 2121, Cyprus \\  
  NIC, DESY, Platanenallee 6, D-15738 Zeuthen, Germany
}

\ead{alexand@ucy.ac.cy}

\begin{abstract}
We discuss the status of  current dynamical lattice QCD simulations in connection to the emerging results  on the strange and charmed baryon spectrum, excited states of the nucleon and the investigation of the structure of scalar mesons. 
\end{abstract}

\vspace*{-1cm}

\section{Introduction}
Simulations of lattice QCD are nowadays being performed with dynamical quarks with masses close to their physical ones. Such simulations with physical pions remove the need for a chiral
extrapolation, thereby eliminating a significant source of a systematic uncertainty that has proved difficult to quantify in the past. Various fermion discretization schemes 
are being employed by various collaborations. MILC has recently presented
results on the pseudoscalar decay constants using Highly Improved Staggered Quark (HISQ) ensembles with the  strange and charm quarks fixed to their physical values and for a range of masses for the two light quarks ($N_f=2+1+1$)   including  physical  values of the light sea-quark masses~\cite{Bazavov:2014wgs}. The BMW collaboration has  produced results on the pion sector using $N_f=2+1$ ${\cal O} (a)$-improved Clover simulations employing HEX smeared links with light quark masses over a range of masses even below the physical pion mass for four lattice spacings~\cite{Durr:2013goa}. A number
of other collaborations are using improved Wilson fermions to simulate with physical or near physical values  of the two dynamical light quark masses, in some cases  including a dynamical strange quark with mass fixed to its physical value. Clover gauge  configurations have been produced by the QCDSF  and PACS-CS collaborations and pion mass $m_\pi\sim 150$~MeV for $N_f=2$~\cite{Bali:2012qs} and $N_f=2+1$ with re-weighing to reach the physical pion value~\cite{Aoki:2009ix}. The European Twisted Mass Collaboration (ETMC) has also generated 
$N_f=2$ gauge configurations using twisted mass fermions including the clover 
term~\cite{Abdel-Rehim:2013yaa}.  Using these 'physical ensembles' one has 
now the possibility to study hadron properties directly.

\section{Hadron spectrum}
 The first quantity that we would like to reproduce from lattice QCD
are the masses of the low-lying hadrons. 
These are extracted from Euclidean correlation functions 
   \be
     G(\vec q, t_s) =\sum_{{\vec x}_s} \, e^{-i\vec {x}_s \cdot \vec q}\, 
     \langle J(\vec {x}_s,t_s)J^\dagger(0) \rangle  = \sum_{n=0,\cdots, \infty} A_n e^{-E_n(\vec{q})t_s}
\stackrel{t_s\rightarrow \infty} {\longrightarrow}A_0 e^{-E_0(\vec{q}) t_s}
\label{two-point}
\ee
 in the large Euclidean time limit, after projecting to zero momentum, $\vec{q}=\vec{0}$.

\begin{figure}[h!]
\begin{minipage}{0.49\linewidth}
{\hspace*{-0.3cm}\includegraphics[width=1.1\linewidth]{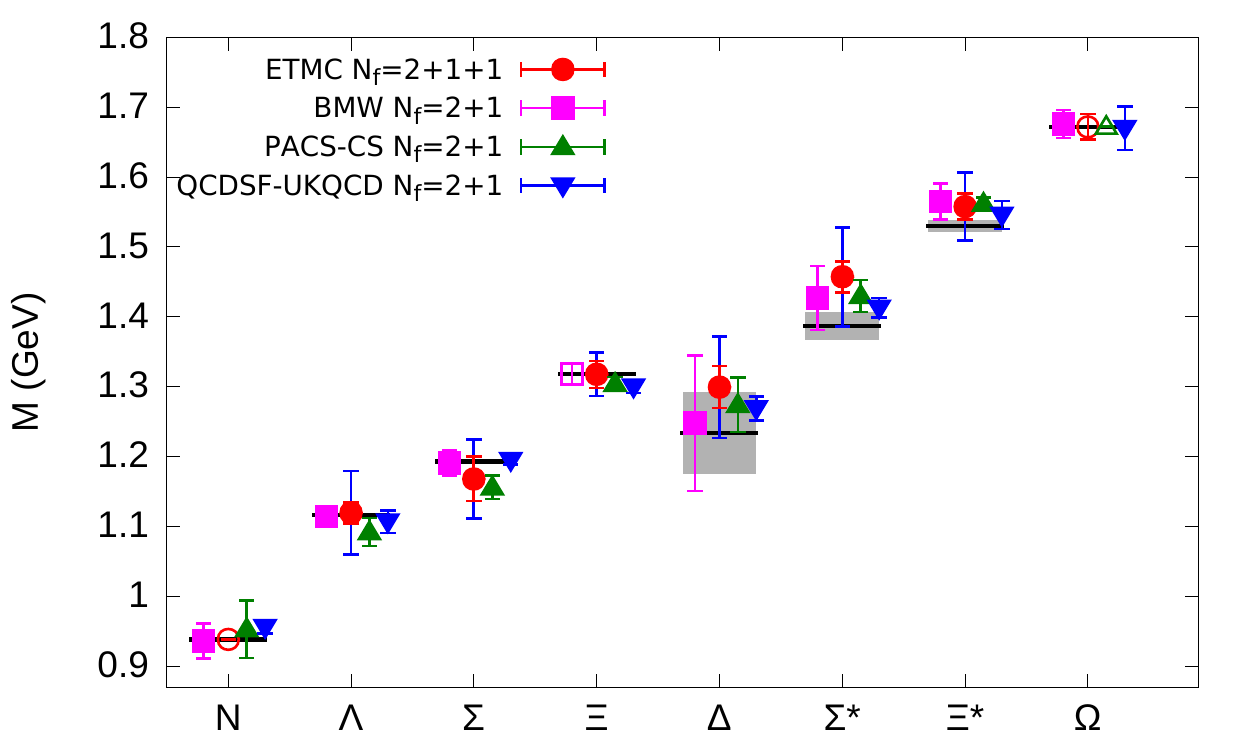}}
\end{minipage}\hfill
\begin{minipage}{0.49\linewidth}
\includegraphics[width=1.1\linewidth]{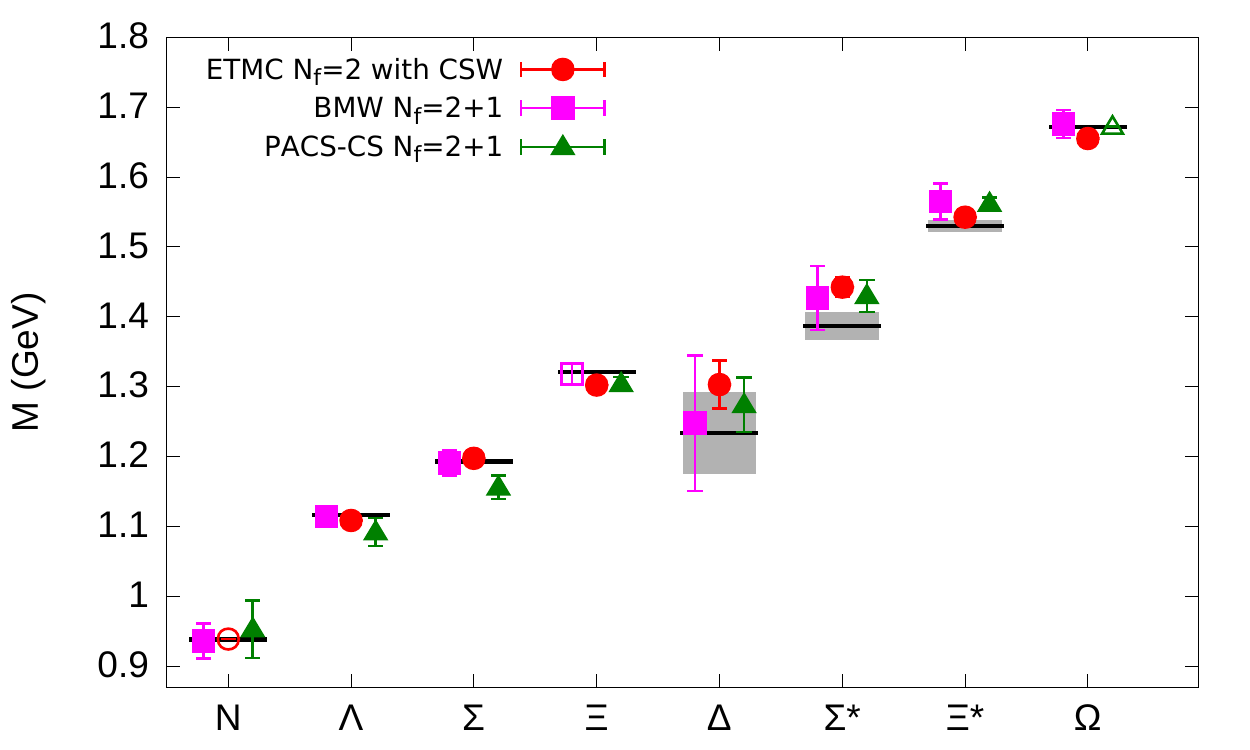}
\end{minipage}
\label{fig:mass}
\caption{Lattice QCD results on the octet and decuplet baryon masses compared to the experimental values shown by the horizontal bands.  Results by ETMC are shown in red circles. Left: using $N_f=2+1+1$ ensembles after performing a chiral extrapolation  (statistical errors are shown in red, whereas the blue error bar includes an estimate of the systematic errors due to the chiral extrapolation~\cite{Alexandrou:2014sha}. Right:
for the physical ensemble~\cite{Alexandrou:2014}. In both plots we also show results using clover fermions from BMW~\cite{Durr:2008zz}  (magenta squares), from PACS-CS~\cite{Aoki:2008sm} (green triangles), and from QCDSF-UKQCD collaborations~\cite{Bietenholz:2011qq} using $N_f=2+1$ SLiNC configurations (blue inverted triangles). Open symbols show  the baryon mass used as input to the calculations.}
\vspace*{-0.3cm}
\end{figure}
 
\begin{figure}[h!]
\begin{minipage}{0.49\linewidth}
\hspace*{-0.3cm}\includegraphics[width=1.1\linewidth]{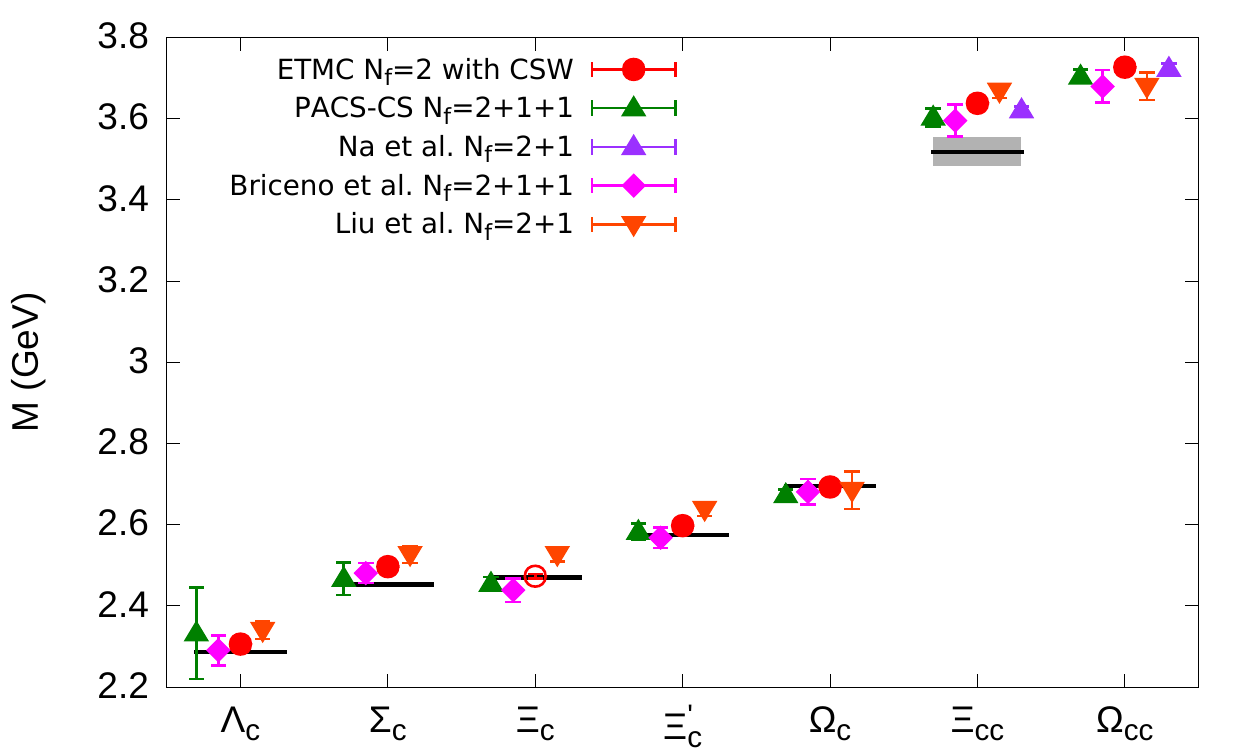}
\end{minipage}\hfill
\begin{minipage}{0.49\linewidth}
\includegraphics[width=1.1\linewidth]{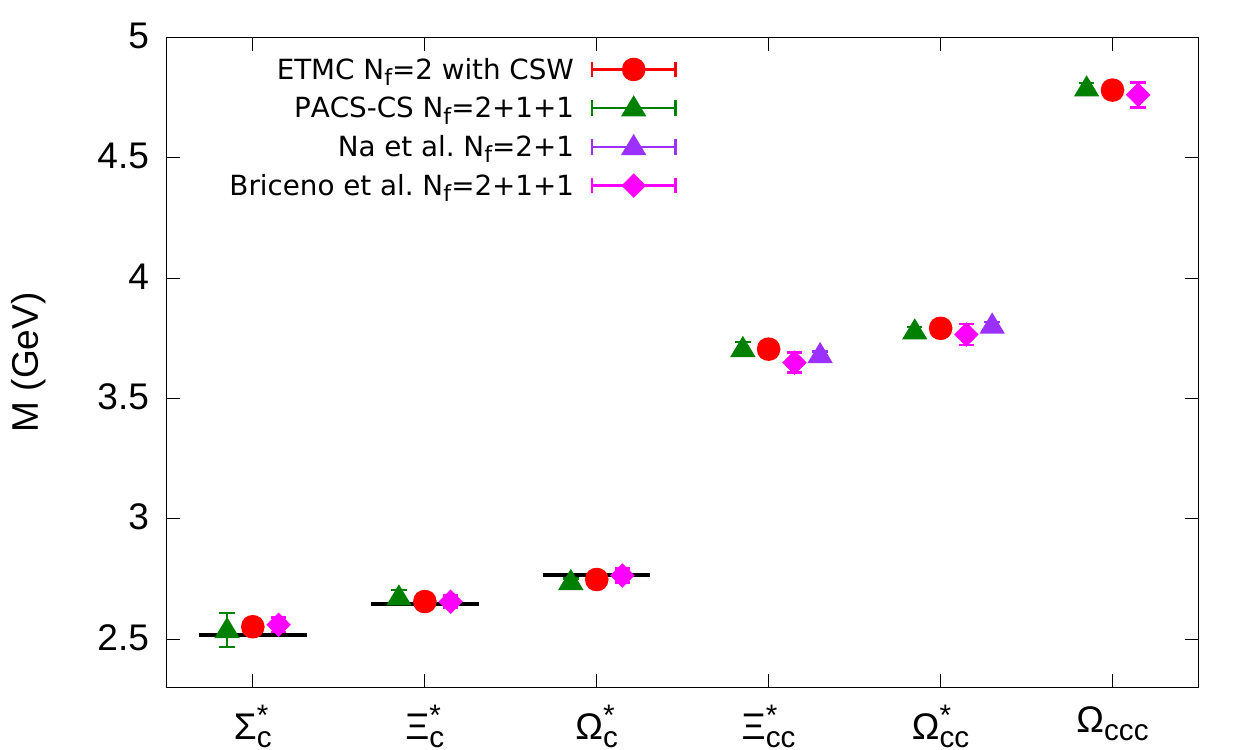}
\end{minipage}
\label{fig:mass charm}
\caption{Results by ETMC are shown in red circles for the mass of the spin-$1/2$ (left) and spin-$3/2$ (right) charmed baryons for the physical ensemble. Included are results from various hybrid actions with staggered sea quarks from Refs.~\cite{Na:2007pv} (purple triangles), \cite{Briceno:2012wt} (magenta diamonds) and \cite{Liu:2009jc} (orange inverted triangles). Results from PACS-CS~\cite{Namekawa:2013vu} are shown in green triangles.}
\vspace*{-0.3cm}
\end{figure}

In Fig.~\ref{fig:mass} we show the spectrum of the octet and decuplet baryons.
We show two sets of results using twisted mass fermions (TMF). One set is obtained with $N_f=2+1+1$ gauge configurations at three lattice spacings, determined using the nucleon mass as  $a=0.094$~fm, $0.082$~fm and $0.065$~fm. 
Thus results can be extrapolated to the continuum limit. The continuum results are chirally extrapolated using heavy baryon chiral perturbation theory to leading and next to leading order. We take the difference between the two orders as an estimate of the systematic error due to the chiral extrapolation, which 
constitutes the  biggest systematic error on the results as can be seen by the blue error bars~\cite{Alexandrou:2014sha}. Results  obtained  at two different lattice  volumes and showed no observable effect within our statistics and thus
volume corrections were not performed. The other set of TMF results shown in Fig.~\ref{fig:mass} is obtained using simulations with physical
values of the light quark masses (physical ensemble), thus requiring no chiral extrapolation, at one lattice spacing and volume.
Both sets agree with experimental values, indicating that finite lattice spacings effects for the physical ensemble are small~\cite{Alexandrou:2014}. These results are now much more precise since the chiral extrapolation was not needed.
 In Fig.~\ref{fig:mass charm} we show the corresponding results for the mass of
the charmed baryons  for the physical ensemble of TMF. As can be seen, the known values of the masses of charmed baryons are reproduced and thus our computation
provides a prediction for the yet unmeasured masses. Our preliminary results for the yet unmeasured mass of the $\Xi_{cc}^*$ is 3.678(8)~GeV, for the  $\Omega^+_{cc}$ is 3.708(10)~GeV, for $\Omega^{*+}_{cc}$ 3.767(11)~GeV and for  $\Omega^{++}_{ccc}$ 4.746(3)~GeV.

\section{Nucleon excited states}
 Having successfully computed the mass of the nucleon and the other low-lying baryons from lattice QCD we turn to the question of the excited nucleon spectrum. We limit our discussion here to the
two lowest states, namely the $P_{11}(1440)$ positive parity resonance known as the Roper and the negative parity state
$S_{11}(1535)$. This is an example where the
 mass  ordering of these states 
is contrary to the prediction
of the constituent quark model where the negative parity state is expected to be lower
in mass than  the $P_{11}$ state.  

 The method to study excited states is to use an enlarged basis of interpolating fields and construct a correlation matrix
\be
G_{jk}(\vec q, t_s)=\sum_{{\vec x}_s} \, e^{-i\vec {x}_s \cdot \vec q}\, 
     \langle{J_j(\vec {x}_s,t_s)J_k^\dagger(0)} \rangle\>, j,k=1,\ldots N \>.
\label{var}
\ee
 Solving the generalized eigenvalue problem (GEVP)
\be G(t)v_n(t;t_0)=\lambda_n(t;t_0) G(t_0)v_n(t;t_0) \rightarrow \lambda_n(t;t_0)=e^{-E_n(t-t_0)}\quad.
\ee
 yields asymptotically the N lowest eigenstates~\cite{Luscher:1990ck}.

We have used a
variational approach  to
study the excited states of the nucleon in the positive and negative
parity channels~\cite{Alexandrou:2013fsu}. Our variational basis consisted of two types of nucleon interpolating fields
with different levels of Gaussian smearings.

\begin{figure}[h!]
\begin{minipage}{0.49\linewidth}
\includegraphics[width=\linewidth,
    keepaspectratio]{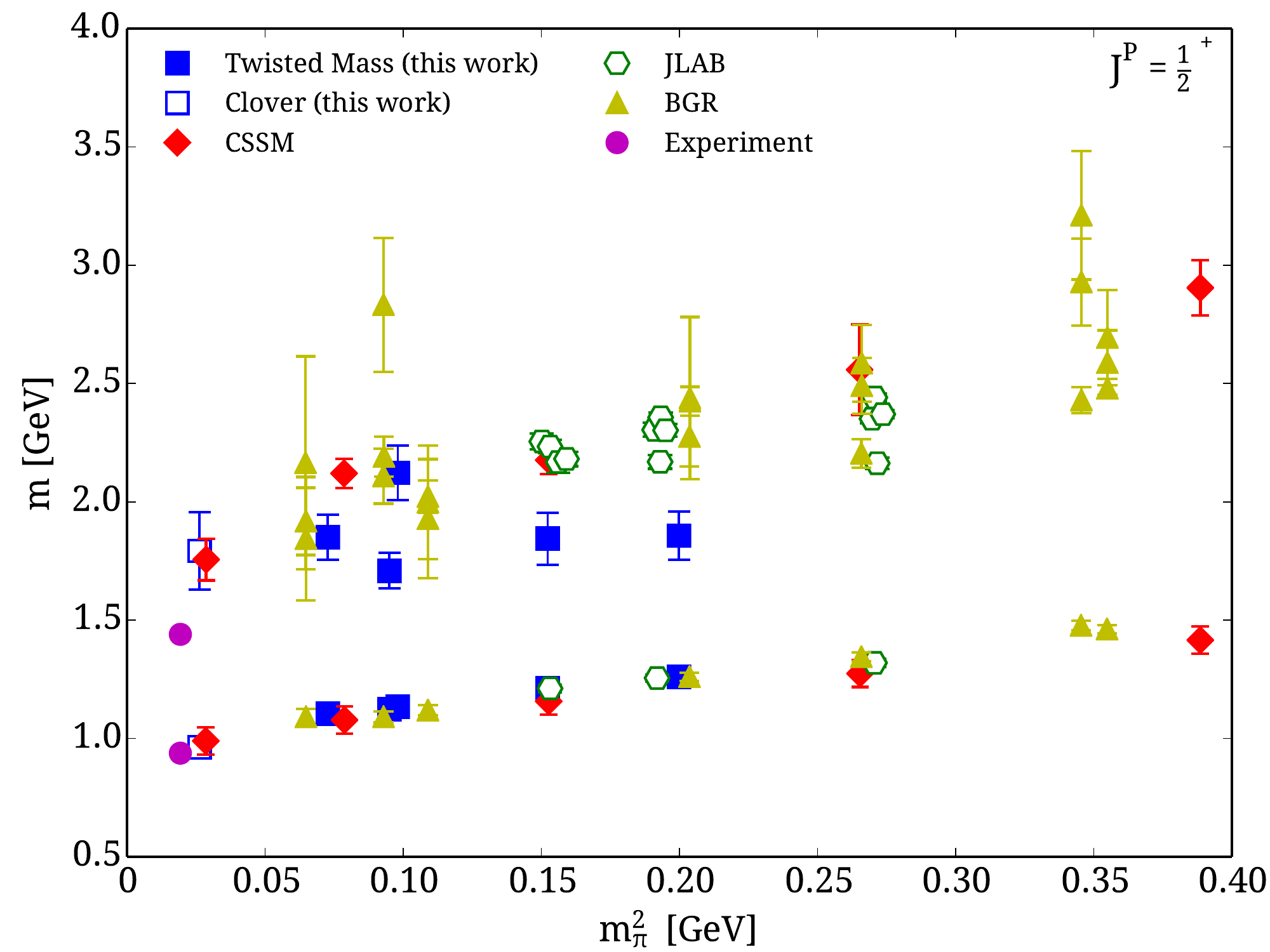}
\end{minipage}\hfill
\begin{minipage}{0.49\linewidth}
  \centering \includegraphics[width=\linewidth,
    keepaspectratio]{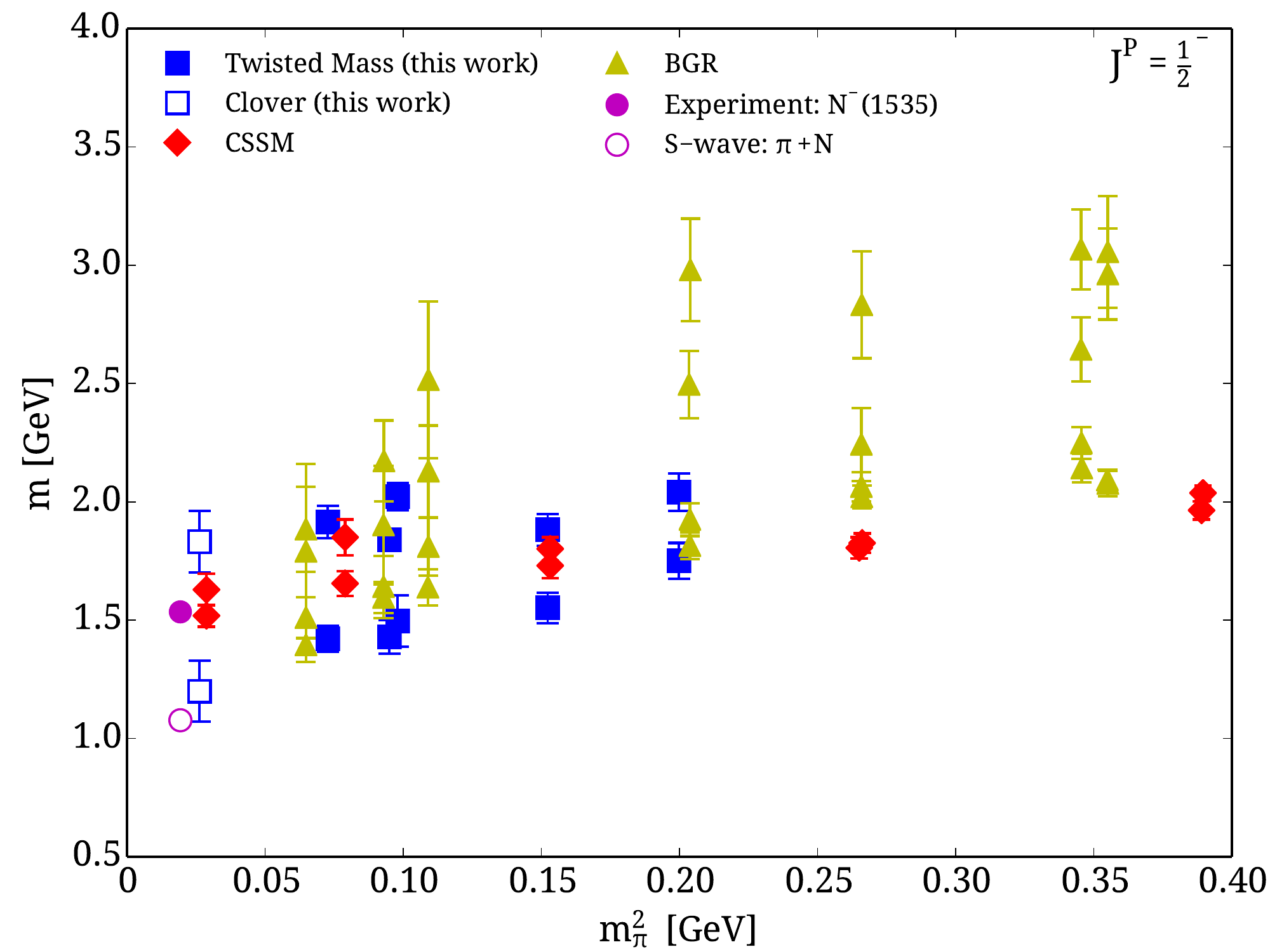}
\end{minipage}
 \caption{\label{compare}  The positive (left) and negative (right) states of the nucleon. Filled and open squares are results using $N_f=2$ twisted mass and clover fermions respectively.   $N_f=2+1$ Clover  fermions  by
    the CSSM collaboration~\cite{Mahbub:2012zz} (red diamonds),
 by the Hadron
    Spectrum Collaboration~\cite{Edwards:2011jj} (open hexagons) 
 using the Chirally Improved Dirac Operator by the
    Bern-Graz-Regensburg (BGR) collaboration~\cite{Engel:2013ig}
    (yellow triangles) are also shown.  
 The CSSM results on the negative parity states are
    from~\cite{Mahbub:2013ala}.}
\end{figure}
\noindent
 We analyzed a total of five ensembles of $N_f=2$ twisted mass fermions
 with pion mass in the range of about 270~MeV to 450~MeV and lattice
 spacing $a=0.089$~fm determined from the nucleon
 mass. For this value of the lattice spacing cut-off effects on the mass of the
 nucleon and hyperons were found to be smaller than the statistical errors.
 Therefore,  we limit ourselves
 to studying only one lattice spacing.  In addition, we analyzed an
 ensemble of $N_f=2$ Clover fermions with pion mass $m_\pi\sim
 160$~MeV and lattice spacing $a\simeq0.073$~fm~\cite{Bali:2012qs}. Our results
are shown in Fig.~\ref{compare}  for the positive and negative channels. Results from other lattice QCD computations are also included.
The first observation is that all lattice results are in reasonable
agreement for the ground state energies of both parity channels. The
second major observation is that our data for the first excited state
of the nucleon in the positive parity channel, although consistent at
near physical pion mass with the other lattice calculation at similar
pion mass, namely that from the CSSM Collaboration, yield a value that is  still higher
than the experimentally measured mass for the Roper. Given that our
lattice volume is comparable to that of Ref.~\cite{Mahbub:2012zz}
volume effects can be responsible for the larger values. In the
negative parity channel we can clearly see that for all pion masses
considered the negative parity ground state is consistent with a
$\pi\,N$ state in an S-wave. To the statistical accuracy available to
us, the first excited negative parity state appears to be converging
to $N^-(1535)$, however the errors are too large to draw concrete
conclusions. It is also apparent that the results on the higher states are
much more spread and carry much larger errors, thus requiring further study.

\section{Scalar mesons}
Using the variational approach one can study scalar mesons.
Our understanding  of the light scalar ($J^P = 0^+$) meson sector below 1 GeV is still unclear. The observed mass ordering of the $f_0(980)$ and $a_0(980)$ states appears inverted from what would be naively expected from the conventional  quark model. 
Instead, the interpretation as four-quark states provides an explanation of the inverted mass values of these scalars~\cite{Pelaez:2014rla}. 
 In this four-quark scenario, the quark content will be given by
\beq
\sigma =ud\bar{u}\bar{d}, & f_0=\frac{1}{\sqrt{2}}\left (us\bar{u}\bar{s}+ds\bar{d}\bar{s}\right), &\nonumber \\
a_0^-=ds\bar{u}\bar{s}, \hspace*{0.2cm} & a_0^0=\frac{1}{\sqrt{2}}\left (us\bar{u}\bar{s}-ds\bar{d}\bar{s}\right), \hspace*{0.2cm} & a_0^+=us\bar{d}\bar{s}\nonumber\\
\kappa^+=ud\bar{d}\bar{s}, \hspace*{0.2cm} &\kappa^0=ud\bar{u}\bar{s},\hspace*{0.2cm} \bar{\kappa}^0=us\bar{u}\bar{d}, \hspace*{0.2cm} &\kappa^-=ds\bar{u}\bar{d}\quad.
\label{4q}
\eeq
 Within this interpretation, the mass degeneracy of $f_0(980)$ and $a_0(980)$ is natural and the mass ordering is understandable. The broad width of the $\sigma$ and $\kappa$ can be also easier explained since the decay channels to $\pi\pi$ and $K\pi$ respectively are OZI super-allowed. 

A number of lattice QCD studies of scalar mesons have been undertaken~\cite{Prelovsek:2008qu,Liu:2008ee,Bernard:2007qf,Wakayama:2012ne}. 
Our on-going investigation of  the low-lying scalar nonet~\cite{Alexandrou:2012rm} focuses first on the study of the  $a_0(980)$ 
 in an effort to shed light on its structure
from QCD. Our variational basis consists of  the conventional two quark operator, as well as, different types of four quark operators, corresponding to two mesons  or a diquark-antidiquark structure. Including   the two quark operator inevitably means that there are diagrams which contain  closed fermion loops, which makes the study even more challenging.
Preliminary results are obtained using a $4\time 4$ correlation matrix
and $N_f=2+1$  clover gauge configurations generated by the PACS-CS 
collaboration~\cite{pacscs_configs}. The lattice size is $32^3\times64$ with lattice 
spacing $a \approx 0.09 \ \rm fm$. Two ensembles are analyzed corresponding to $m_\pi \approx 300$~MeV, and $m_\pi \approx 150$~MeV. Results are still preliminary and currently we are improving our calculations.   
Including explicitly  scattering states  for the $\eta_{ss}+\pi$ and $K+\bar{K}$ type with explicit projection 
 to zero momentum is expected to improve our ability to isolate the $a_0(980)$ 
state.
 
\section{Conclusions}
Simulations at the physical point are now becoming available
enabling the computation of the masses of the  low-lying hadron  without the need
of chiral extrapolation. This eliminates an up to now ill-determined systematic error from lattice QCD evaluations.
 Results on other benchmark quantities such as
the nucleon axial charge  $g_A$, and momentum fractions $\langle x \rangle _{u-d}$  are also being computed directly at the physical point~\cite{Alexandrou:2013jsa, Alexandrou:2014}. We will need high statistics and careful cross-checks to finalize results
on these quantities at the physical point.
The study of excited states and resonances is under way but many challenges 
need still to be resolved.

\vspace*{-0.3cm}

\section{Acknowledgments}
I would like to thank my close collaborators  A. Adbel-Rehim,  J. Berlin, M. Dalla Brida, V. Drach,  M. Gravina, K. Jansen, Ch. Kallidonis, G. Koutsou, T. Leontiou and  M. Wagner for their
valuable contributions without which this work would not have been 
possible. 
I am also indebted to the QCDSF collaboration~\cite{Bali:2012qs} for making the clover configurations available to us.

\end{document}